\documentclass{article}
\usepackage{graphicx}
\usepackage[utf8]{inputenc}

\usepackage{amsmath}
\usepackage{amssymb}
\newcommand{\be}{\begin{equation}}
\newcommand{\ee}{\end{equation}}

\begin{document}

\title{About ionization zones remaining after supernovae explosions}
\author{{G.S. Bisnovatyi-Kogan}\\
	Institute of Applied Mathematics, Acad. of Sci. of the USSR \\
Translated from Astronomicheskii Zhurnal, Vol. 49, No. 2, \\
pp. 453-454, March-April, 1972 \\
Original article submitted September 16, 1971}

\maketitle

\begin{abstract}
\noindent It is shown that investigation of ionization zones left after supernova explosions can yield information on the mechanism active in those explosions.
\end{abstract}

The Stromgren remnant zones can be used to determine the mode of radiation that gave rise to them, in view of the very long recombination time in a rarefied medium [1]

\be
\tau_r=\frac{4\cdot 10^4}{n_e}\left(\frac{T_e}{10^3\, K}\right)^{1/2}\,\,\, {\mbox {years}}.
\label{eq1}
\ee

\noindent These zones can be observed around rapidly evolving stars [2] or quasars [3]. This article will show that the investigation of Stromgren remnant zones around supernovae occurring in the galaxy can yield  information on the color temperature of the supernova radiation and can be helpful in making a definite choice between competing theories accounting for the light-curves of supernovae.

At the present time, emission occurring in supernova explosions outside of any dependence on the explosion mechanism is accounted for either in terms of the emergence of a shock wave into the extended envelope [4], or in terms of decay of radioactive elements [5]. The explanation invoking helium fluorescence [6] runs up against the formidable difficulties pointed out in [4].

Since the temperature at light maximum from the equilibrium radiation is $\sim 3 \cdot 10^{4}$ K in the model used in [4], and $6  \cdot 10^{3}$ K in the model used in [5], investigation of zones found in supernovae that exploded in the local galaxy over the past $\sim 10^{4}$ years would be helpful in making a choice between those mechanisms in the formation of the light-curve. If the radius of the Stromgren zone $R_{s}$ is determined by the total ionizing radiation of the supernova $\varepsilon_i$, then we have

\be
R_s=\left(\frac{3\varepsilon_i}{4\pi n_H I_H}\right)^{1/3}= 33\left(\frac{\varepsilon_i/10^{50}}{ n_H }\right)^{1/3}\,\,\, {\mbox{pc}}.
\label{eq2}
\ee
 In our estimates we neglect ionization of the helium, and consider only hydrogen. The mass of the ionized medium is then

\be
M_i=\frac{\varepsilon_i m_p}{I_n}=3.8\cdot 10^3\,(\varepsilon_i/10^{50})\, M_\odot.
\label{eq3}
\ee
When $e_i >10^{49}$ ergs, we have $M_i >380 M_\odot$, which is much greater than the mass of the exploding star, so that ionization of the matter flowing out of the star can be ignored, and we need consider only the interstellar medium. In the case of Planck radiation, the ratio of the radii of the ionization zones for different temperatures is proportional to the cube root of the ratio $(\varepsilon_{i1}/\varepsilon_{i2})$, which is $(\varepsilon_{i1}/\varepsilon_{i2})$ = $10^{8}$ when $T_1 = 3 \cdot 10^{4}$ K and $T_{2}=6\cdot 10^3$ K at the same total luminosity.

Consequently, the ionization zones are comparable to the Stromgren zones of hot stars [7] when the energies of supernovae in ionizing radiation are from $\sim 10^{49}$ to $10^{51}$ ergs.

In discussing the remnants of explosions of supernovae of age $10^{3}$ - $10^{4}$ years in our Galaxy, we can state that recombination does not have time to go to completion at $n_e\sim 1$. We estimate the radius to which the shock wave following in the wake of the light wave propagates. This radius is

\be
R_s\approx 10\left(\frac{t}{10^3\,{\mbox{years}}}\right)\left(\frac{V}{10^4\,{\mbox{km/s}}}\right)\,\, {\mbox{pc}}.
\label{eq4}
\ee
Consequently, a remnant ionization zone with dimensions greater than those of the zone traversed by the shock wave (the supernova remnant proper) is possible in the case of galactic supemovae.

The following information can be obtained from observations of Stromgren remnant zones.

\medskip

 {\bf 1}. Ionization zone is smaller than the supernova remnant. In this case we can immediately state the upper bound of the supernova's ultraviolet emission. This is the case with the Crab Nebula, in which no ionization zone is observed. Using the dimension of the nebula
    $\sim 1 pc = 3\cdot 10^{18 }$ cm, and Eq.(\ref{eq2}), we can obtain

\be
\varepsilon_{i}\le 7\cdot 10^{44} ergs. 
\label{eq5}
\ee
Recalling that the light energy released in the explosion has been estimated [8] at $10^{49}$ ergs, and
$\langle n_H\rangle \approx 0.16$ follows from the dispersion measure of pulsars [9], we find the brightness temperature was not greater than $10^{4}$ K  in the case of the supernova explosion in the Crab Nebula, which argues in favor of radioactive decay [5], rather than a shock wave in the extended envelope.

\medskip 

 {\bf 2}. {Ionization zone is very large
$\gg 100$ pc, much larger than the supernova remnant.} 
In this case the ionizing radiation must have been enormous, according to Eq.(\ref{eq2}). Near the remnant of the supernova Vela X we find an ionized nebula of large dimension (cylinder of radius 400 pc, height 100 pc), which has been ascribed [10, 11] to the action of the supernova explosion. But that conclusion cannot be accepted as definitive, since the radius of the ionization zone
of the three hot stars located within that nebula is 450 pc, given the density $\langle n_e^2\rangle^{1/2} = 0.16$ cm$^{-3}$, rather than the 144 pc assumed in [10] for $\langle n_e^2\rangle^{1/2} = 1$. This means that the origin of that zone can be explained entirely by the existence of those hot stars, when we consider the uncertainty in the problem.

\medskip

{\bf {3}}. The existence of an intermediary ionization zone $\lesssim 100$ pc does not introduce any unambiguity into the interpretation, since it can be accounted for in terms of the ionizing radiation from the O-star, which was fully capable of being a presupemova up to the point of explosion.

An ionization zone of dimension $R_{s}$ = 5.7 pc at $n_e = 15$ cm$^{-3}$ and $I = 18$ eV has been discovered [12] in the remnant of the supernova CasA. This ionization zone could have been formed by a presupernova, if it was a star of the class of B0 hot stars.
The value of $10^{50}$ ergs reported in [12] can be regarded on that account as the upper bound of the ultraviolet emission in the explosion of Cas A.

I take this opportunity to express my gratitude to Ya. B. Zel'dovich, V. S. Imshennik, and R. A. Sunyaev for helpful discussions.


\begin{thebibliography}{99}

\bibitem{1}  L. Spitzer, Diffuse Matter in Space, Interscience (1968).

\bibitem{2}  G. S. Bisnovatyi-Kogan and R. A. Sunyaev, Astron. Zh., 47, 441 (1970) [Sov. Astron. -AJ, 14, 351 (1970)].

\bibitem{3}  J. Arons and R. McKee, Astrophys. J. Letters, 5, 127 (1970).

\bibitem{4}  E. K. Grasberg, V. S. Imshennik, and D. K. Nadezhin, Astrophys. Space Sci., 10, 3 (1971).

\bibitem{5}  S. A. Colgate and Ch. McKee, Astrophys. J., 157, 623 (1969).

\bibitem{6}  P. Morrison and L. Sartori, Astrophys. J., 541 (1969).

\bibitem{7}  V. S. Sobolev, Treatise on Theoretical Astrophysics [in Russian], Nauka (1967).

\bibitem{8}  I. S. Shklovskii, Supernova Stars [in Russian], Nauka (1966).

\bibitem{9}  R. J. Gold, Astrophys. Space Sci., 10, 265 (1971).

\bibitem{10}  J. C. Brandt, T. P. Stecher, D. L. Crawford, and S. P. Haran, Astrophys. J. Letters, 163,
99 (1971).

\bibitem{11}  W. H. Tucker, Astrophys. J. Letters, 167, L85 (1971).

\bibitem{12}  M. Peimbert and S. van den Bergh, Astrophys. J., 167, 223 (1971).

\end{thebibliography}
\end{document}